\documentclass[11pt,letter,defpaper]{article}
\usepackage{psfig}
\usepackage{defpaper}
\usepackage{latexsym}
\include{psfig}

\eatreminders

\def\papernumber #1 raised #2 {
\vspace{-#2}
\vbox to 0pt{\hfill\framebox{\bf Paper Number #1}}
\vspace{#2}
}

\newcommand{\usepdf}[1]{}
\newcommand{\ifpdf}[1]{}
\newcommand{\useps}[1]{#1}
\newcommand{\ifps}[1]{#1}

\newcommand{\deltaplus}[1]{\delta^{+}_{#1}}
\newcommand{\deltaminus}[1]{\delta^{-}_{#1}}
\newcommand{\oldval}[1]{#1^{old}}

\newcommand{\unionsp}{~~\union~~}
\newcommand{\diffchildren}{\mbox{\it diffChildren}}
\newcommand{\fullchildren}{\mbox{\it fullChildren}}
\newcommand{\diffcost}{\mbox{\it diffCost}}
\newcommand{\totaldiffcost}{\mbox{\it totalDiffCost}}
\newcommand{\localdiffcost}{\mbox{\it localDiffCost}}
\newcommand{\differential}{\delta}

\title{
Materialized View Selection and Maintenance Using Multi-Query 
Optimization\thanks{Work
partly supported by a Govt. of India, Department of Science and 
Technology Grant, and by an IBM University Partnership Program Grant.  
The work of Prasan Roy was supported by an IBM Research Fellowship.
Ramamritham was also supported in part by NSF grant IRI-9619588.
}}
\author{{\bf Hoshi Mistry}$^1$ 
\and {\bf Prasan Roy}$^1$
\and {\bf Krithi Ramamritham}$^{1,2}$ 
\and {\bf S. Sudarshan}$^1$ \\
$^1$ - Indian Institute of Technology, Bombay, India\\
     $^2$ - Univ. of Massachusetts, Amherst \\
     \{hoshi,prasan,krithi,sudarsha\}@cse.iitb.ernet.in \\
}
\date{}

\begin{document}

\eat{
{\Large
\hspace*{\fill} Paper Number: {\bf 194} \\
\hspace*{\fill} Category: {\bf Research} \\
}
\\
{\large
\begin{center}
{\Large\bf Materialized View Selection and Maintenance Using Multi-Query 
		Optimization} \\
~\\
{\bf Hoshi Mistry}, {\bf Prasan Roy},  {\bf Krithi Ramamritham} and
{\bf S. Sudarshan}\\
\end{center}
~\\
\begin{tabular}{ll}
Contact author &  Krithi Ramamritham \\
& krithi@cs.umass.edu\\
& Computer Science Department \\
& University of Massachusetts \\
& Amherst, Mass. 01003-4610 \\
& (413) 545-0196 (office) \\
& (413) 545-1249 (fax) \\
\end{tabular}
}
}

\thispagestyle{empty}
\pagebreak
\setcounter{page}{1}

\maketitle
\begin{abstract}
Because the presence of views enhances query performance, materialized
views are increasingly being supported by commercial database/data
warehouse systems.  Whenever the data warehouse is updated, the
materialized views must also be updated. However, whereas the amount
of data entering a warehouse, the query loads, and the need to obtain
up-to-date responses are all increasing, the time window available for
making the warehouse up-to-date is shrinking. These trends necessitate
efficient techniques for the maintenance of materialized views.

In this paper, we show how to find an efficient plan for maintenance
of a {\em set} of views, by exploiting common subexpressions
between different view maintenance expressions.  These common subexpressions
may be materialized temporarily during view maintenance.
Our algorithms also choose subexpressions/indices to be materialized
permanently (and maintained along with other materialized views),
to speed up view maintenance.
While there has been much work on view maintenance in the past,
our novel contributions lie in exploiting a recently developed framework
for multiquery optimization to efficiently find good view maintenance
plans as above.
In addition to faster view maintenance, our algorithms
can also be used to efficiently select materialized views to speed up
workloads containing queries.

\end{abstract}

\sections{Introduction}

Materialization of views can help speed up query and update
processing. Views are especially attractive in data warehousing
environments because of the query intensive nature of data
warehouses. However, when a warehouse is updated, the materialized
views must also be updated.  Typically, updates are accumulated and
then applied to a data warehouse.  Loading of updates and view
maintenance in warehouses has traditionally been done at night.  While
the need to provide up-to-date responses to an increasing query load
is growing and the amount of data that gets added to data warehouses
has been increasing, the time window available for making the
warehouse up-to-date has been shrinking.  These trends call for
efficient techniques for maintaining the materialized views as and
when the warehouse is updated.

Given multiple views, the view maintenance problem can be seen as
computing the expressions corresponding to the ``delta'' of the views,
given the ``delta''s of the base relations that are used to define the
views. The contributions of this paper lie in the exploitation of the
Multi-Query Optimization (MQO) framework along with our recently
developed efficient algorithms for MQO, to compute the delta
expressions corresponding to the multiple views defined in a data
warehouse. 

It is not difficult to motivate that query optimization techniques are
important for choosing an efficient plan for maintaining a view.  For
example, consider the expression $(A \Join B) \Join C$, where $A$, $B$
and $C$ are multisets (i.e., relations with duplicates).  Given that
the multiset of tuples $\deltaplus{C}$ is inserted into $C$, the
change to the view is given by $(A \Join B) \Join \deltaplus{C}$.
This expression can equivalently be computed as $(A \Join
\deltaplus{C}) \Join B $ and by $(B \Join \deltaplus{C}) \Join A $,
one of which may be substantially cheaper to compute.  Further, in
some cases the view may be best maintained by recomputing it, rather
than by finding the differentials as above.  
Vista \cite{vista98:viewmaint:opt}
describes how to extend the Volcano query optimizer \cite{gra:vol} to
choose the best plan for computing the differential of the result of
an expression.

Given a set of queries, multiquery optimization \cite{tim:mul}
provides the possibility of reducing costs by computing shared
subexpressions once, materializing them temporarily, and reusing them
where required in the given set of queries.  Although multiquery
optimization was earlier viewed as expensive, our recent work 
\cite{rssb00:mqo} has provided efficient algorithms for multiquery
optimization, making it practical.  In this paper, we provide
practical solutions to the problem of optimizing the update
of a set of materialized views, by exploiting these algorithms.

Specifically, our contributions are as follows.
\begin{enumerate}
\item We extend the multiquery optimization algorithms to find
the best plan for computing the differential of a set of expressions,
by exploiting shared subexpressions.

Sharing of subexpressions occurs when multiple views are being
maintained, since related views may share subexpressions, and as a
result the maintenance expressions may also be shared.  Furthermore,
sharing can occur even within the plan for maintaining a single view,
as we illustrate later in the paper.
 
Our algorithms choose shared expressions to be temporarily materialized
during view maintenance,  and choose view maintenance plans that
utilize these temporarily materialized results.


\item 
Just as the presence of views allows queries to be evaluated more
efficiently, the maintenance of these views can be made more efficient
by the presence of additional views/indices \cite{rss96:matview}. 
That is, given a set of materialized views to be maintained,
we need to choose what additional indices and views should be materialized
to minimize overall view maintenance costs. 

The choice of additional views must be done in conjunction with
choosing plans for maintaining the views.
For instance, a plan that seems quite inefficient could become the best
plan if some intermediate result of the plan is chosen to be materialized
and maintained.

Our contribution here is to show how to extend the multiquery optimization
algorithms of \cite{rssb00:mqo} to tackle the problems of
selecting permanent materialized views, in conjunction with choosing 
the best plans for updating the views.
\end{enumerate}
We show how to cleanly integrate the choice of 
expressions/indices to be {\em permanently} materialized, with the choice 
of expressions/indices to be {\em temporarily} materialized.


It is worth pointing out that although our focus in this paper is to
speed up view maintenance, our algorithms can also be used to
choose extra temporary and permanent views in order to speed up a
workload containing queries and updates (that trigger view
maintenance).

There has been much earlier work on choosing a set of views to be materialized
and maintained to optimize given workloads of queries and
updates. The major differences between our work and earlier work can be
summarized as follows (we outline the differences in detail in Section~\ref{sec:relwork}):
\begin{enumerate}
\item
Given a set of related materialized 
views, temporarily materializing common subexpressions could have significant
benefit. However,  earlier work did not consider how to
exploit common subexpressions by temporarily materializing them
because of their focus on permanent materialization and 
common subexpressions involving differential relations 
cannot be permanently materialized.  

\item The earlier work does not cover efficient techniques for the
implementation of materialized view selection algorithms, in
particular, their integration with query optimizers.  In the context
of materialized view maintenance, this is an
important problem since the cost of view maintenance can be reduced by
the presence of (additional) indices on relations, and of appropriate extra
materialized views.

In contrast, we show how to efficiently choose views/indices to 
be (permanently) materialized by extending the multiquery optimization 
algorithms of \cite{rssb00:mqo}.
\end{enumerate}

The rest of the paper is organized as follows.
We outline related work in Section~\ref{sec:relwork},
and provide the reader with some background in view maintenance in
Section~\ref{sec:background}.
We describe the DAG structure used to represent queries in 
Section~\ref{sec:dagrep}, and algorithms to find optimal update
plans (without materializing additional views) in Section~\ref{sec:opt}.
Section~\ref{sec:greedy} presents an optimized greedy algorithm for 
selecting extra views for materialization.
Section~\ref{sec:perf} outlines results of a performance
study, and Section~\ref{sec:concl} concludes the paper.

\sections{Related Work}
\label{sec:relwork}

There has been a large volume of research on incremental view maintenance
in the past decade.
Amongst the early work on computing the differential results of 
operations/expressions was Blakeley et al. \cite{blakeley86:viewmaint}.
More recent work in this area includes
\cite{griffin95:viewmaint,colby96:viewmaint,mumick97:viewmaint}.
Gupta and Mumick \cite{gm95:viewsurvey} provide a survey 
of view maintenance techniques.

Blakeley et al.~\cite{blakeley86:viewmaint} and 
Ross et al. \cite{rss96:matview} noted that the computation of the
expression differentials has the potential for benefiting from
multiquery optimization.
In the past, multiquery optimization was viewed as too expensive for
practical use.
As a result they did not go beyond stating that 
multiquery optimization could be useful for view maintenance.
Our recent work in \cite{rssb00:mqo} provides efficient 
heuristic algorithms for multiquery optimization, and demonstrates that
multiquery optimization is feasible and effective.

There has been much work on selection of views to be materialized.
One notable early work in this area was by Roussopolous \cite{rou82:view}.
Ross et al. \cite{rss96:matview} considered the selection of 
extra materialized views to optimize maintenance of other 
materialized views/assertions, and mention some heuristics.
Labio et al. \cite{lqa97:phys} provide further heuristics.
The problem of materialized view selection for data cubes has seen
much work, such as \cite{venky:sigmod96}, who propose a greedy heuristic for 
the problem.  Gupta \cite{gupta97:viewsel} and Gupta and Mumick
\cite{gupta99:viewmaint} extend some of these ideas to a wider class 
of queries.    
However, (a) none of the above papers consider implementation details that are 
important for efficient selection of views, and
(b) none of these consider how to optimize view maintenance expressions.

Vista \cite{vista98:viewmaint:opt} describes how to extend the 
Volcano query optimizer
to optimize view maintenance.  However, she does not consider the
materialization of expressions, whether temporary or permanent.
Optimizations that exploit knowledge of foreign key dependencies 
can be used to detect that certain join results involving 
differentials will be empty \cite{qgmw96:views,vista98:viewmaint:opt}.

Roy et al. \cite{rssb00:mqo} consider
how to perform multiquery optimization by selecting subexpressions/indices for 
temporary materialization.  They present important optimizations of a greedy 
heuristic for materialized view selection that makes the heuristic 
practical.  
However, they do not consider updates or view maintenance, which 
is the focus of this paper.
We utilize the optimizations proposed in \cite{rssb00:mqo}, 
but the extensions required to to take update costs into account, 
and optimize view maintenance expressions, are non-trivial.

There has been earlier work on multiquery optimization, 
including \cite{tim:mul,kyu:imp,sg:tkde90} and more recently
\cite{shivku98:transview}, but none of these consider updates.

\sections{Background and Motivation}
\label{sec:background}


We assume that updates (inserts/deletes) to relations are logged in 
corresponding {\em delta} relations, which are made available to the
view refresh mechanism.  We assume for each relation $r$,
there are two relations $\deltaplus{r}$ and $\deltaminus{r}$ denoting,
respectively, the (multiset of) tuples inserted into and deleted from 
the relation $r$.

The view refresh mechanism is invoked as part of an update transaction 
for immediate update, or periodically for deferred updates.  
In the second case, updates performed by many transactions may be 
collected together in the {\em delta} relations $\deltaplus{r}$ and
$\deltaminus{r}$ for a relation $r$.
Our techniques work regardless of whether the updates are immediate
or deferred.

\subsections{Computing the Differential of an Operation}
\label{sec:op:diff}

There is a considerable amount of literature on computing differentials
of operations, as outlined in Section~\ref{sec:relwork}.
For completeness, we briefly review techniques for computing 
the differential of an operation in the multiset relational algebra.

\subsubsections{Differentials of Joins}

Consider a multiset join $ A \Join B$, and suppose $A$ and $B$
are updated by inserting the multisets of tuples $\deltaplus{A}$ and
$\deltaplus{B}$ respectively.
Let $\oldval{A}$ and $\oldval{B}$ refer to the old values of
$A$ and $B$, that is their contents before the update.
The multiset of tuples that get added to the view $V$ are denoted by
$\deltaplus{V}$, and can be computed as: 
\[ \deltaplus{V}  = (\deltaplus{A}\Join \oldval{B}) \union 
		    (\oldval{A} \Join \deltaplus{B}) \union 
                    (\deltaplus{A} \Join \deltaplus{B}) \]
View $V$ is then updated as follows: $ V \leftarrow V \union \deltaplus{V}$\\

Similarly, if tuples $\deltaminus{A}$ and $\deltaminus{B}$ are 
deleted from $A$ and $B$ respectively, the multiset of tuples
\[ \deltaminus{V}  = (\deltaminus{A}\Join \oldval{B}) \union 
		     (\oldval{A} \Join \deltaminus{B}) \union 
                     (\deltaminus{A} \Join \deltaminus{B}) \]
get deleted from $V$, which is then updated by:
$ V \leftarrow V  - \deltaminus{V}$.

Updates can be modeled as deletes followed by inserts.
If both inserts and deletes are present on a relation, 
we get a more complex expression for updating the relation.
\[ V \leftarrow V  \union (\oldval{A} \Join \deltaplus{B})
		   \union (\deltaplus{A} \Join \oldval{B})
		  \union  (\deltaplus{A} \Join \deltaplus{B})
		  -       (\oldval{A} \Join \deltaminus{B})
		  -       (\deltaminus{A} \Join \oldval{B})
		  \union  (\deltaminus{A} \Join \deltaminus{B})
\]

In contrast if only one input, say $A$, is updated by only insertion
the change in the view is much easier to compute: 
\[ \deltaplus{V} =  \deltaplus{A} \Join \oldval{B} \]
and similarly for deletions on $A$, 
\[ \deltaminus{V} =  \deltaminus{A} \Join \oldval{B} \]

To keep expressions simple (and for another reason which we 
describe later in Section~\ref{sec:queryopt:motivation})
we assume that updates are propagated
one relation at a time, and only one type of update at a time.
This simply means we compute the effect of all inserts on $A$,
then update $A$ with $\deltaplus{A}$, then compute the effects of
all deletes on $A$, update $A$ with $\deltaminus{A}$.
We then proceed with inserts to $B$, and then with deletes from $B$.

The net result is the same as if the more complicated expressions
are used, but the expressions we need to deal with are much simpler.
Note that an operation may have two complex expressions as inputs,
and if both use a particular relation, even with this restriction
there may be differential results on both its inputs, in which
case the more complex expression allowing differentials on both
inputs must be used.

\subsubsections{Differentials of Other Operations}

If the result of a groupby/aggregate operation, such as 
$_{A}{\cal G}_{count(B)}(E)$, has been materialized (and the
aggregate function is distributive), the change in 
the aggregate result can be computed using only the 
changes ($\deltaplus{E}$ and $\deltaminus{E}$) to the input $E$,
and the old result of the aggregation.\footnote{For some operations
like $average$, the count of tuples in each group
must also be materialized.  Even for the $sum$ operation, the
count of tuples is needed to deal with deletions.}
The group-by/aggregation operation is executed on the tuples from the 
delta relations, and the results are merged into the existing
materialized view using a merge operation.
For more details, see, e.g., \cite{gm95:viewsurvey},
and for extensions to operations such as median, 
see \cite{rsss94:aggcons}.

To compute the differential result of an aggregate/grouping operation
whose result has not been materialized, we would have to recompute the
aggregate values for all groups which are affected by the update.
This may involve significant extra work.\footnote{There are techniques, 
such as \cite{colby96:viewmaint,mumick97:viewmaint}, that use differentials 
of more complex forms, such as changes in the value of an aggregate result,
to avoid recomputing aggregate values in some special cases.
Our techniques can be extended to deal with such differentials, but 
for simplicity we do not consider such techniques here.}

Standard techniques are available for computing the differentials
of other operations, such as duplicate elimination (and projection),
and outer joins \cite{griffin95:viewmaint,gupta97:outerjoin}.
We omit details but note that we can use these techniques without any 
changes to our optimization algorithms.

\subsections{Computing the Differential of an Expression}

Views are defined by potentially complex expressions, hence
we need to find the differential of an entire expressions.

\subsubsections{Generating a Differential Expression}

Techniques, such as that of \cite{griffin95:viewmaint} can be
used to generate an expression that computes the differential
of a given expression.  However, the resultant expression
can be very large -- exponential in the size of the query.
For instance consider the view $V = A \Join B \Join C$,
with inserts on all three relations.
The differential in the result of $V$ can be computed as
\begin{quotation}
\noindent
$ (\deltaplus{A} \Join B \Join C) \unionsp
  (A \Join \deltaplus{B} \Join C) \unionsp
  (A \Join B \Join \deltaplus{c}) \unionsp  
  (A \Join \deltaplus{B} \Join \deltaplus{C}) \unionsp $ \\
$ (\deltaplus{A} \Join B \Join \deltaplus{C}) \unionsp
  (\deltaplus{A} \Join \deltaplus{B} \Join C) \unionsp 
  (\deltaplus{A} \Join \deltaplus{B} \Join \deltaplus{C}) $ 
\end{quotation}
The size of this expression is exponential in the number of relations.
Optimizing such large expressions can be quite expensive, since 
query optimization is exponential in the size of the expression.
There are many common subexpressions in the above expression, and
the above expression could be simplified by factoring, to get:\\
\hspace*{1in}
$ (\deltaplus{A} \Join B \Join C) \unionsp 
  ((A \union \deltaplus{A}) \Join \deltaplus{B} \Join C)  \unionsp 
  ((A \union \deltaplus{A}) \Join (B \union \deltaplus{B}) \Join
          \deltaplus{C})  $ \\
But creating simplified forms of differential expressions is 
difficult with more complex expressions containing operations
other than join.

Therefore our algorithms use an alternative technique, which 
we outline in the next section.

\subsubsection{Propagating Updates Up An Expression}
\label{sec:diffprop:expression}

An alternative to generating a differential expression is to
propagate differentials up an expression \cite{rou82:view,rss96:matview}.
Propagation is best understood by visualizing an expression as a tree.
The differential of a node in the tree is computed using
the differential (and if necessary, the old value) of its
inputs, as described earlier.
We start at the leaves of the tree, and proceed upwards, computing
the differential expressions at each node.

For example, consider an expression $(A \Join B) \Join C)$, 
and suppose we wish to propagate inserts to $A$.
\reminder{Draw a figure for this expression}
We can do so by first computing the differential of the node $A \Join B$
as $\deltaplus{A} \Join B$.
We then join this differential with $C$, which is the other input 
of its parent node, to get the differential of the parent.

As mentioned earlier, if there are updates to multiple relations, 
we propagate one type of update to one relation at a time.  
Doing so simplifies the expressions for computing the differentials,
as outlined in Section~\ref{sec:op:diff}, and permits a different 
evaluation plan to be chosen for each expression; this is
essential for efficient view maintenance, as we will see next, in
Section~\ref{sec:queryopt:motivation}.


The process of computing the differential of an
expression can be expressed purely in terms of how to compute 
the differentials for each operation in the expression.
There is no need to rewrite the entire expression.
Note also that the procedure for computing differentials of an expression
can be easily extended to handle expressions using new types of operations, 
so long as we have a way of incrementally computing the differential
of the operation.

\subsubsection{The Role of Query Optimization}
\label{sec:queryopt:motivation}
 
Consider an expression $ A \Join (B \Join C)$, 
and suppose tuples are inserted into $A$.
We can compute the differential of the result as 
$ \deltaplus{A} \Join (B \Join C)$.
If we compute this expression as shown, we would need to compute 
$B \Join C$, which does not involve any $\delta$ relation, and 
hence may be large and expensive.
A better way of evaluating the differential 
may be $ (\deltaplus{A} \Join B) \Join C$.
Note that the two variants are logically equivalent.  

Thus, for efficient differential computation, query optimization 
must be applied to the update expressions to choose the cheapest
variant, as proposed in \cite{vista98:viewmaint:opt}.

Furthermore, note that if we wish to compute the differential
when tuples are inserted into $C$, the plan
$ (\deltaplus{C} \Join B) \Join A$ or 
$ (\deltaplus{C} \Join A) \Join B$ may be preferable to
$ (A \Join B) \Join \deltaplus{C}$.
Thus, using a single expression, such as $(A \Join B) \Join C$  to
propagate differentials to $A$, $B$ and $C$ is likely to perform badly for
at least one of the differentials.

For this reason, we propagate differentials of only one relation
at a time, and choose a separate plan for each differential propagation. 

We use a query optimizer for choosing best plans for such propagation,
and our optimizer uses a DAG representation that compactly represents
all expressions equivalent to a given expression.
Since all alternative expressions above are available, the best one
can be chosen for the propagation of each differential.
We present details in Sections~\ref{sec:dagrep} and \ref{sec:opt}.


Note also that recomputation of a materialized view is always an alternative
to computing the differential in its result and updating it. 
Thus, the query optimizer must choose recomputation over incremental
view maintenance, if recomputation is cheaper.

\subsections{The Role of Multi-Query Optimization in View Update}
\label{sec:MQO}

Multi-query optimization attempts at exploiting common sub-expressions 
within a query, or across queries in a batch of queries submitted together,
to reduce the query evaluation cost. 
In the context of view update, sharing can occur across the tasks of 
computing differentials of different views, 
or even within the task of computing the differential
of a single view, as we show below. 


It is easy to see that related queries may share subexpressions,
and if so, it may be best to compute the shared subexpression once,
materialize it, and reuse it.
However, this decision must be done in a cost based manner,
as the following example from \cite{rssb00:mqo} illustrates.

\begin{example}
\label{example:motivating}
Let $Q_1$ and $Q_2$ be two queries
whose locally optimal plans (i.e., individual best plans) are
$(R \Join S) \Join P$ and $(R \Join T) \Join S$ respectively.
The best plans for $Q_1$ and $Q_2$ do not have any
common sub-expressions.
However, if we choose the alternative plan $(R \Join S) \Join T$
(which may not be locally optimal) for $Q_2$,
then, it is clear that $R \Join S$ is a common sub-expression
and can be computed once and used in both queries.
This alternative with sharing of $R \Join S$ may be the globally
optimal choice.

On the other hand, blindly using a common
sub-expression may not always lead to a globally optimal strategy.
For example, there may be cases where the cost of joining the expression
$R \Join S$ with $T$ is very large compared to the cost
of the plan $(R \Join T) \Join S$; in such cases it may make no
sense to reuse $R \Join S$ even if it were available.
\end{example}

In the context of view maintenance, if two materialized views
have common subexpressions, as in the example above, 
the expressions for computing the differential of the 
common subexpression would also be shared.


To illustrate subexpression sharing  possibilities within a single
view maintenance query, consider a view $V$ defined as in the example below.
\begin{example}
\label{ex:diff}
Let view $V = A \Join B \Join C \Join D$.
Suppose there are inserts on all four relations.

%
%

The differential of $V$ can then be computed using
\begin{quotation}
\noindent
$ (\deltaplus{A} \Join B \Join C \Join D) \unionsp 
  ((A \union \deltaplus{A}) \Join \deltaplus{B} \Join C  \Join D) \unionsp $\\
$  ((A \union \deltaplus{A}) \Join (B \union \deltaplus{B}) \Join
          \deltaplus{C} \Join D) \unionsp 
  ((A \union \deltaplus{A}) \Join (B \union \deltaplus{B}) \Join
          (C \union \deltaplus{C}) \Join \deltaplus{D}) $ 
\end{quotation}
The above expression represents algebraically the effect of
propagating differentials one at a time, as described in 
Section~\ref{sec:diffprop:expression}.

Note that there are several potential common subexpressions
in the above expression.
For instance, if the plans chosen for the first two terms of the 
above union are
$ (\deltaplus{A} \Join (B \Join (C \Join D))) $ and
$ ((A \union \deltaplus{A}) \Join (\deltaplus{B} \Join (C  \Join D))) $,
then $C \Join D$ is a common subexpression of the two.
If the above plans are chosen, the subexpression can be computed once
and shared.
Similarly, $ (A \union \deltaplus{A}) \Join (B \union \deltaplus{B})$
is potentially a common subexpression.

The alternative plans of  
$ (((\deltaplus{A} \Join B) \Join C) \Join D) $ and
$ ((((A \union \deltaplus{A}) \Join \deltaplus{B}) \Join C)  \Join D) $
offer no sharing possibilities, but may still be cheaper. 
\end{example}

Which the above plans should be used, and whether the common
subexpressions should materialized and shared is a decision for the 
multiquery optimizer to make in a cost-based manner.
Thus, it is the job of the multiquery optimizer to find the best overall plan
taking sharing possibilities into account.



\sections{DAG Representation of Queries}
\label{sec:dagrep}

Our algorithms use an extended form of the DAG representation of queries
used, for instance, in Volcano \cite{gra:vol}.
In this section we summarize the DAG representation and 
terminology from \cite{rssb00:mqo}.

An AND--OR DAG is a directed acyclic graph whose nodes can be divided
into AND-nodes and OR-nodes;  the AND-nodes have only OR-nodes 
as children and OR-nodes have only AND-nodes as children. 

An AND-node in the AND-OR DAG corresponds to an algebraic operation, 
such as the join operation ($\Join$) or a select operation ($\sigma$).
It represents the expression defined by the operation and its inputs. 
Hereafter, we refer to the AND-nodes as {\em operation nodes}.
An OR-node in the AND-OR DAG represents a set of logical 
expressions that generate the same result set; the set of such expressions
is defined by the children AND nodes of the OR node, and their inputs.
We shall refer to the OR-nodes as {\em equivalence nodes} henceforth.

\begin{figure*}
\centerline{
\ifpdf{
\mbox{\pdfimage width 6.0in {figures/dagex.pdf} \relax}
}
\ifps{
\psfig{file=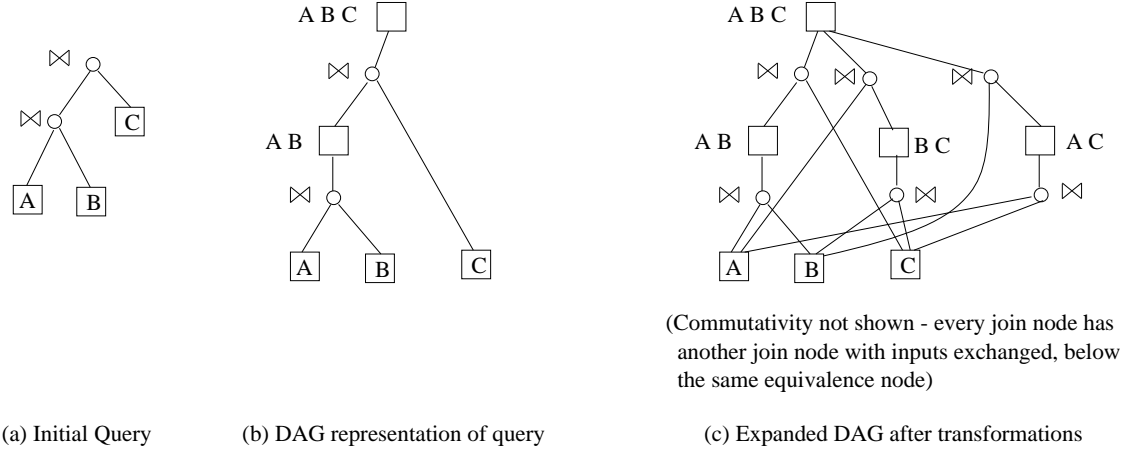,width=6in}
}
}
\caption{Initial Query and DAG Representations}
\label{fig:vol:dag}
\end{figure*}

\subsections{Representing a Single Query}

A given query is initially represented directly in the AND-OR DAG 
formulation.
For example, the query tree of Figure~\ref{fig:vol:dag}(a) is initially
represented in the AND-OR DAG formulation, as shown 
in Figure~\ref{fig:vol:dag}(b).
Equivalence nodes (OR-nodes) are shown as boxes, while
operation nodes (AND-nodes) are shown as circles.

The initial AND-OR DAG is then expanded by applying all possible 
transformations on every node of the initial query DAG representing
the given set of queries. 
Suppose the only transformations possible are join associativity
and commutativity.
Then the plans $A \Join (B \Join C)$ and $(A \Join C) \Join B$,
as well as several plans equivalent to these modulo commutativity
can be obtained by transformations on the initial AND-OR-DAG
of Figure~\ref{fig:vol:dag}(b).
These are represented in the DAG shown in Figure~\ref{fig:vol:dag}(c).
We shall refer to the DAG after all transformations have been applied
as the {\em expanded DAG}.
Note that the expanded DAG has exactly one equivalence node for every 
subset of $\{A, B, C\}$; the node represents all ways of computing 
the joins of the relations in that subset.  

\subsections{Representing Sets of Queries in a DAG}
\label{sec:dagrep:queryset}

Queries are inserted into the DAG structure one at a time.
When a query is inserted, equivalence nodes and operation nodes are created
for each of the operations in its initial query tree. 
Some of the subexpressions of the initial query tree may be equivalent to
expressions already in the DAG.  Further, subexpressions of a query 
may be equivalent to each other, even if syntactically different.
For example, query may contain a subexpression that is logically
equivalent to, but syntactically different from another subexpression
of the query (e.g., $(A \Join B) \Join C$, and $A \Join (B \Join C)$).

Before the second subexpression is expanded, the DAG would contain 
two different equivalence nodes representing the two subexpressions.  
\cite{rssb00:mqo} modifies the Volcano DAG generation algorithm such that 
whenever it finds nodes to be equivalent (in the above example, after applying 
join associativity) it {\em unifies} the nodes, replacing them by 
a single equivalence node.
The Volcano optimizer \cite{gra:vol} already has a hashing-based scheme to 
efficiently detect repeated expressions, thereby avoiding creation of 
new nodes that are equivalent to existing nodes.  
The extension of \cite{rssb00:mqo} additionally unifies existing 
logically equivalent nodes.
Another extension is to detect and handle {\em subsumption} derivations.
For example, $\sigma_{A<5}(E)$ can be computed from
$\sigma_{A<10}(E)$ if they both appear in a set of queries.
Similarly, if we have aggregations $_{dno} {\cal G}_{sum(Sal)} (E) $ and
$_{age} {\cal G}_{sum(Sal)} (E) $, we can introduce a new equivalence
node $_{dno,age} {\cal G}_{sum(Sal)} (E) $ and add derivations of
the other two from this one.
\eat{
}
For more details of unification and subsumption derivations involving
selections and aggregation, see \cite{rssb00:mqo}.

\subsections{Physical Properties}
\label{ssec:phys:dag}

It is straightforward to refine the above AND-OR DAG representation to
represent {\em physical properties} \cite{gra:vol}, such as sort order,
that do not form part of the logical data model, and obtain 
a physical AND-OR DAG
\footnote{For example, an equivalence node is refined to multiple
physical equivalence nodes, one per required physical property, in the
physical AND-OR DAG.}.  
The presence of an index on a result is also modeled as a physical property
of the result by \cite{rssb00:mqo}, making the code that handles
physical properties also perform index selection.
Physical properties of intermediate results
are important; for example, if an intermediate result is sorted on a
join attribute, the join cost can potentially be reduced by using a
merge join.  This also holds true of intermediate results that are
materialized and shared.  
Our implementation indeed handles physical properties, including
sort orders and indices, but to keep the description simple we do not 
explicitly consider physical properties.  



\sections{Finding Optimal Plans}
\label{sec:opt}

We first outline how to find optimal plans for queries,
following \cite{rssb00:mqo}, and then outline
extensions to find optimal plans for view maintenance.
In both cases, we assume that the set of views chosen for
materialization is fixed.
In Section~\ref{sec:greedy} we outline how to integrate the
choice of views to materialize with the choice of optimal
plans for view maintenance.

\subsections{Finding Optimal Plans for Queries}
\label{sec:costeq:basic}

The Volcano optimization algorithm finds the best plan for each node
of the expanded DAG by performing a depth first traversal of the DAG.
Costs are defined for operation and equivalence nodes.
The computation cost of an operation node is $o$ is defined as follows:\\
\hspace*{1in}
$compcost(o) = $ cost of executing $(o)$ + $\Sigma_{e_i \in children(o)}
compcost(e_i)$\\
The children of $o$ (if any) are equivalence nodes.
The computation cost of an equivalence node $e$ is given as \\
\hspace*{1in}
$compcost(e) = min \{ compcost(o_i) | o_i \in children(e) \} $\\
and is $0$ if the node has no children (i.e., it represents a 
relation).\footnote{Relation scans are explicitly represented as an 
operation and assigned a cost.}
Note that the
cost of executing an operation $o$ also takes into account the cost
of reading the inputs, if they are not pipelined.

Volcano also caches the best plan it finds for each equivalence node,
in case the node is re-visited during the
depth first search of the DAG.

A simple extension of the Volcano algorithm to find best plans given
a set of materialized views is described in \cite{rssb00:mqo}.  
We outline this extension below.

Let $reusecost(m)$ denote the cost of reusing the materialized
result of $m$, and let $M$ denote the set of materialized nodes.

To find the cost of a node given a set of nodes $M$ have been materialized,
we simply use the Volcano cost formulae above for the query, with the
following change.
When computing the cost of an operation node $o$, if an input equivalence
node $e$ is materialized (i.e., in $M$),  the minimum of
$reusecost(e)$ and $compcost(e)$ is used for computing $compcost(o)$.
Thus, we use the following expression instead:\\
\hspace*{1in}
$compcost(o, M) = $ cost of executing $(o)$ + $\Sigma_{e_i \in children(o)}
  C(e_i, M)$
\vspace{-0.1in}
\begin{tabbing}
xxxxxxxxxxxxxxxxxxx\=where $C(e_i, M)$\= \kill
\>where $C(e_i, M) = compcost(e_i)$ if $e_i \not\in M$ \\
\>\>$=min(compcost(e_i, M), reusecost(e_i))$ if $e_i \in M$.\\
\end{tabbing}
\vspace{-0.3in}
and $compcost$ for equivalence nodes is defined as before.
Thus, the extended optimizer computes best plans for the query in
the presence of materialized results.  The extra optimization overhead
is quite small.

\subsections{Extending the DAG Structure for Computing Differentials}
\label{sec:dag:diff}


We now outline how to extend the DAG structure to 
represent the differentials of a set of expressions.
We first construct the expanded DAG for the given expression (or set of
expressions). 
As in Volcano, each equivalence node in the DAG has a set of logical
properties such as the schema of the expression result, and
estimated statistics about the result such as number of tuples.
Once the best plan is computed for a node, it is cached in case
it is needed later during optimization.

If there are $n$ relations, $R_1, \ldots, R_n$,
we need to store information about the differentials of the
node with respect to 
$\deltaplus{R_1}$, $\deltaminus{R_1}$, 
$\deltaplus{R_2}$, $\deltaminus{R_2}$, 
and so on until 
$\deltaplus{R_n}$, $\deltaminus{R_n}$. 
We number these updates as $1, \ldots, 2n$, and use these numbers to
identify the update.


To optimize differential plans, each equivalence node 
stores information for the differentials of the expression 
with respect to each update type, in addition to information about
the full result.
Each equivalence node $e$ therefore stores an array of 
$2n$ records, as below.
Each odd numbered entry $2i-1, i=1..n$,  of this array contains:
\begin{enumerate}
\item logical properties (such as schema and estimated statistics) 
of the differential of $e$ with respect to inserts on $R_i$
\item the best plan for computing the differential of $e$ with respect
to inserts on $R_i$ 
\item the logical properties of the full result of the equivalence
node after inserts and deletes to relations $R_1, \ldots, R_{i-1}$
have been propagated
\end{enumerate}
Similarly, each even numbered entry $2i, i=1..n$,  of this array contains
similar information on differentials and best plans with respect 
to the deletes on $R_i$, and the logical properties for the full result
of the equivalence node after inserts and deletes to relations 
$R_1, \ldots, R_{i-1}$, and inserts to $R_i$ have been propagated.

In addition, as in the original representation, each node
stores the best plan for (and cost of) recomputing the entire result of the
node after all updates have been made on the base relations.

The logical properties of the differentials are computed
by a bottom-up traversal of the DAG.
We describe later how the best plans for computing differentials
are computed and stored.  If an equivalence node does not 
depend on relation $R_i$, we flag this during the above bottom-up traversal,
and set the plans in entry $2i$ and $2i+1$ to be null.

The traversal also computes and stores an estimate of the execution cost of 
the differential version of each operation in the DAG 
(such as a join or an aggregation).  The properties of 
the differentials of its inputs, as well as the full
version of the input, where required, are used to compute this estimate.




\subsections{Finding Optimal Plans for Updates}
\label{sec:costeq:diff}

We now outline how to find optimal plans for updates,
using the above mentioned DAG representation.
Recall the example from Section~\ref{sec:queryopt:motivation},
with the expression being $ A \Join (B \Join C)$, and an insert on $A$,
the plan $ (\deltaplus{A} \Join B) \Join C$, is likely to be more 
efficient than $ \deltaplus{A} \Join (B \Join C)$, and should be considered.
Luckily for us, the DAG representation of the query represents
$ (A \Join B) \Join C$ in addition to $ A \Join (B \Join C)$
(see Figure~\ref{fig:vol:dag}).

We now extend the technique for finding optimal plans
for queries described in Section~\ref{sec:costeq:basic}, 
to find the optimal way of 
propagating the differential $\deltaplus{A}$.

Some equivalence nodes do not depend on some relations, and their 
differential with respect to the relation will be empty.
Let $\diffchildren(o, i)$ denote all equivalence node children of $o$ whose
differential is non-empty on update $i$, and 
$\fullchildren(o, i)$ denotes all children of $o$ whose
full results are required to compute the differential of $o$,
in conjunction with $\diffchildren(o, i)$.

For instance, $\diffchildren$ for an operation that joins
$A$ with $ (B \Join C)$, with respect to an insert on $B$,
is the node $B \Join C$, and correspondingly $\fullchildren$
of the node is $A$.

Given an operation node $o$ in the DAG, let $\differential(o, i)$ denote 
the differential of operation $o$ with respect to update $i$.
Also let $\localdiffcost(o, i)$ denote the cost of executing the 
operations in $\differential(o, i)$, without counting the cost of 
generating its inputs.

Similarly, for an equivalence node $e$, let $\differential(e, i)$ denote
the differential result of $e$ with respect to update $i$.
Then, the total cost of generating the differential result of an
operation node $o$ with respect to update $i$, $\diffcost(o, i)$ can be 
computed by:
\[
	\localdiffcost(o, i) + 
	\Sigma_{e_j \in \diffchildren(o, i)} \diffcost(e_j, i) + 
	\Sigma_{e_j \in \fullchildren(o, i)} compcost(e_j)
\]
The cost of computing the differential of an equivalence node $e$ 
with respect to update $i$ is given as \\
\hspace*{1in}
$\diffcost(e, i) = min \{ \diffcost(o_j, i) | o_j \in children(e) \} $\\
and is $0$ if the node has no children (i.e., it represents a relation or
a relation differential).
The definition of $compcost$ is as defined earlier in 
Section~\ref{sec:costeq:basic}, and represents the cost of 
recomputation of the node after updates have been performed on the
database relations.

\reminder{Should we multiple count compcost for different differentials?  
I think its OK, but factoring could provide sharing without materialization
which we are not considering.  We should perhaps mention this difference
between factoring and our materialization costing explicitly somewhere.}

The above formula is extended for the case where some
nodes are materialized, as follows.
Note that the full result of a node may be materialized, and independently,
any of its differential results may also be (temporarily) materialized.
Let the set of materialized results be $M$; 
For an operation node $o$, we compute $ \diffcost(o, M, i) $ as:
\[
\localdiffcost(o, i) + 
                   \Sigma_{e_j \in \diffchildren(o, i)} C(e_j, M, i) +
                   \Sigma_{e_j \in \fullchildren(o, i)} C(e_j, M)
\]
where $C$ is defined as follows:
if $\differential(e, i)$ is not materialized (i.e., not in $M$), \\
\hspace*{1in}$ C(e, M, i) = min \{ \diffcost(o_j, M, i) | o_j \in 
		children(e) \} $\\
and if $\differential(e, i)$ is materialized (i.e., in $M$),\\
\hspace*{1in}$ C(e, M, i) = min ( reusecost(e, i), 
            min \{ \diffcost(o_j, M, i) | o_j \in children(e) \} ) $ \\
and $reusecost(e, i)$ denotes the cost of reusing the materialized
result of $\differential(e, i)$.
Also, $C(e_j, M)$ plays the same role for the full result of node $e_j$,
as defined in Section~\ref{sec:costeq:basic}.

For an equivalence node $e$, 
$\diffcost(e, M, i) = min \{ \diffcost(o_j, M, i) | o_j \in children(e) \} $\\
and is $0$ if the node has no children (i.e., it represents a relation or
a relation differential).
That is, $\diffcost$ represents the cost of computing the differential,
even if the differential is materialized.

For each equivalence node $e$, the operation node corresponding to the 
minimum cost in the above formula defines the (top node of the) best 
plan for $\differential(e, i)$, given that results in 
$M$ are materialized.

Further, we can compute the total cost of computing the differential
of a node as \\
\hspace*{1in}$\totaldiffcost(e,M) = \Sigma_{i=1\ldots2n}\diffcost(e,M)$.


\reminder{Need an example to illustrate above}


We perform a single traversal of the DAG to compute the costs
for each equivalence/operation node, based on the above equations.
During the traversal we also cache the best (minimum cost) plan 
computed for each differential, just as we cache the best plans 
for each full result.

Note that if both inputs to a join $ E_1 \Join E_2$ are expressions using 
a common relation $R$, an update to $R$ results in changes to 
both inputs, and as a result, the update expression for the join 
is 
$ (\deltaplus{E_1} \Join E_2) \union ((E_1 \union \deltaplus{E_1}) \Join
\deltaplus{E_2}) $.
In this case, a join in the original expression has been converted
into a union of two joins.  The best plan for each join is found, giving
the best plan for the entire expression, and this combined best plan
must be stored (and used to compute the cost of finding the differential).



Optimizations that exploit knowledge of foreign key dependencies 
can be used to detect that certain join results involving 
differentials will be empty \cite{qgmw96:views,vista98:viewmaint:opt}.
For instance, if $r.B$ is a foreign key into $s.A$, then the join of
$\deltaplus{s}$ and $r$ will be empty.
Based on this, parts of the differential expression can be
detected to be empty, and eliminated during optimization.


\sections{The Greedy Algorithm for Selecting Materialized Views}
\label{sec:greedy}

Till now we assumed that the set of materialized nodes is fixed.
We now describe how to integrate the choice of extra materialized
views/indices with the choice of best plans for view maintenance.
Our algorithm is based on a greedy heuristic.
We first present the basic algorithm, then some optimizations,
and extensions, below.

\subsections{The Basic Greedy Algorithm}
\label{sec:greedy:basic}

As outlined earlier, we first take the given set of materialized 
views ${\cal V}$, and build a DAG structure on the expressions defining 
the views.
The nodes of the DAG corresponding to views in ${\cal V}$ are
marked as already chosen for materialization.

We consider both full and differential results for materialization.
A result is identified by a node and an update number
(in our implementation a full result is identified by the update number
$0$, and differential results by numbers $1\ldots2n$).


If a result is chosen for temporary materialization, 
we must take into account the cost of computing it.
And if it is chosen for permanent materialization, we must take
into account the cost of maintaining it (we need not consider the cost
of initial materialization since it is a one time cost).

The cost of maintaining a node incrementally is the sum of the
costs of its differentials:\\
\hspace*{1in}$ maintcost(n,M) = \totaldiffcost(n,M) + mergeCost(n) $ \\
where $mergeCost(n)$ denotes the cost of updating the
materialized result of $n$ using the differentials.

For a full result $n$, we define \\
\hspace*{1in} $cost(n, M) = min(compcost(n,M)+matcost(n), maintcost(n, M))$\\
where $matcost(n)$ denotes the cost of writing out the computed result of
$n$.
That is, when finding the cost of the full result of a materialized node,
we take the minimum of the cost via recomputation and the cost via
computing the differentials.

For a differential result $x = \differential(n, i)$, we define \\
\hspace*{1in}$cost(x, M) = \diffcost(n,M, i) + matcost(x)$.\\
Given a set $S$ of results (full/differential),
let $cost(S, M)$ be defined as \\
\hspace*{1in} $ cost(S, M) = \Sigma_{q \in S}  cost(q,M) $ \\
Given a set of results $M$ already chosen to be materialized, and a 
result $x$, the benefit of additionally materializing $x$, $benefit(x, M)$, 
is defined as:\\
\hspace*{1in} 
$ benefit(x, M) = cost(M, M) - (cost(M, \{x\} \cup M) + cost(x, M)) $ \\
Note that $(cost(M, \{x\} \cup M) + cost(x, M)) $ is equivalent to
$cost(M \cup \{x\}, M \cup \{x\})$.


\begin{figure}
\begin{small}
\ordinalg{
Procedure {\sc Greedy} \\
{\em Input:} \> \> Expanded DAG for ${\cal V}$,  the initial set of
materialized views,
\\
\> \> and the set of candidate equivalence nodes (and their differentials) for 
materialization \\
{\em Output:} \> \> Set of nodes/differentials to materialized \\
\> X = ${\cal V}$ \\
\> Y = set of candidates equivalence nodes/differentials for materialization  \\
\> while (Y $\neq \phi$) \\
\>\>  Pick the node $x $ with the highest $ benefit(x, X)$ \\ 
\>\>  if ($benefit(x,X) < 0$) \\
\>\>\>  break; /* No further benefits to be had, stop */ \\
\>\>Y  = Y - x;~~~X = X $\cup$ \{x\} \\
\> return X
}
\end{small}
\vspace{-2mm}
\caption{The Greedy Algorithm}
\label{fig:greedy}
\end{figure}

Figure~\ref{fig:greedy} outlines a greedy algorithm 
that iteratively picks nodes to be materialized.
The procedure takes as input the set of candidate results 
(equivalence nodes, and their differentials) for materialization.
At each iteration,
the node $x$  that gives the maximum reduction in the cost 
if it is materialized, is chosen to be added to $X$.


The procedure not only selects results for maintenance, but also
decides on how they should be maintained.
Specifically, for full results, it chooses the cheaper of recomputation 
(including the cost of storing the result), and differential
computation (including the cost of performing the computed differential
updates).
If recomputation is cheaper for a result, and the result was not
part of the given set of materialized view, the result can be materialized
temporarily during view maintenance, and discarded later.
Differential results that are chosen to be materialized are
materialized only temporarily since they are only used during 
view maintenance.


%

\subsections{Optimizations}
\label{sec:greedy:opt}

The greedy algorithm as described above can be expensive due to 
the large number of times the function $benefit$ is called,
(which in turn calls the expensive function $cost()$).

Some important optimizations to the greedy algorithm for multi-query
optimization are presented in \cite{rssb00:mqo}.
We use two of the optimizations, with some extensions for 
handling differentials: 
\begin{enumerate}
\item There are many calls to benefit (and thereby to
$cost()$) at line L1 of Figure~\ref{fig:greedy}, with different parameters.
A simple option is to process each call to $cost$ independent 
of other calls. 
However, observe that the set of materialized nodes which is the second
argument of $cost$ changes minimally in successive calls ---
successive calls take parameters of the form $cost(R, \{x\} \cup X)$, 
where only $x$ varies.
That is, instead of considering $x_1 \cup X $ for materialization,
we are now considering storing $x_2 \cup X$ for materialization.
The best plans computed earlier does not change for nodes that are
not ancestors of either $x_1$ or $x_2$.
It makes sense for a call to leverage the work done by a previous call 
by recomputing best plans only for ancestors of $x_1$ and $x_2$.

A novel incremental cost update algorithm is presented in \cite{rssb00:mqo}.
This algorithm maintains the state of the DAG (which includes previously 
computed best plans for the equivalence nodes) across calls  
to $cost$, and may even avoid visiting many of the ancestors of 
$x_1$ (which is unmaterialized) and $x_2$ (which is materialized).

In our context of finding update plans, we have to modify the incremental
cost update algorithm slightly.
\begin{enumerate}
\item 
If the full result of a node is materialized, 
we update not only the cost of computing the full result of 
each ancestor node, but also the costs for the $2n$ differentials
of each ancestor node since the full result may be used in any of the
$2n$ differentials.
Propagation up from an ancestor node can be stopped
if there is no change in cost to computing the full result or any of
the differentials.
\item 
If the differential of a node with respect to update $i$ is materialized,
we update only the differentials of its ancestors with respect to update $i$.
Propagation can stop on ancestors whose differentials with respect to
$i$ do not change in cost.
\end{enumerate}

\item The monotonicity optimization works as follows.  With the 
greedy algorithm as presented above, in each iteration the benefit 
of every candidate node that is not yet materialized is recomputed
since it may have changed.

The monotonicity optimization is based on the assumption that the benefit
of a node cannot increase as other nodes are chosen to be materialized -- while 
this is not always true, it is often true in practice.  
The monotonicity optimization makes the above assumption, and does not 
recompute the benefit of a node $x$ if the new benefit of some node 
$y$ is higher than the previously computed benefit of $x$.  
It is clearly preferable to materialize $y$ at this stage, rather 
than $x$ --- assuming monotonicity holds, the benefit of 
$ x$ could not have increased since it was last computed,
and it cannot be the node with highest benefit now, hence its
benefit need not be recomputed now.  

Thus, recomputations of benefit are greatly reduced.

%
\end{enumerate}

\cite{rssb00:mqo} presents a third optimization based on
potential sharability of nodes between queries.
This optimization is not relevant here, since even a node that is
not sharable may be worth materializing permanently.

For the purpose of this paper, the details of the above optimizations
are {\em not} critical, but the interested reader may refer to
\cite{rssb00:mqo} for details.


The Greedy procedure can be extended in a straightforward manner
to consider a workload of queries, along with periodic updates, 
and to choose the best set of results (and indices) to materialize, 
to minimize the cost of the queries and view update.  
Some optimizations are needed to handle large workloads \cite{rssb00:mqo}.
We can also introduce limits on space for storing permanently 
materialized results and temporarily materialized results.
Results can then be materialized in the order of benefit 
per unit space, instead of just benefit.

\sections{Performance Study}
\label{sec:perf}

We implemented the algorithm described earlier for finding optimal
plans for view maintenance.
Like the existing multiquery optimization code, the new code
implements index selection along with selection of results to
materialize.
Our current implementation has a restriction in 
that it only considers full results for materialization, although a
version which also considers differential results for materialization 
should be ready shortly.  Thus our estimated benefits are actually
conservative, and we may be able to get even better results once
the full implementation is ready.  However, the benefits are already
very significant.
\vspace*{-1ex}
\subsection{Performance Model}
\vspace*{-2ex}

We used a benchmark consisting of TPC-D queries (and some variants 
based on the same TPC-D schema).  
The performance measure is {\em  estimated execution cost}, called
{\em plan cost} in the performance graphs.  
Our cost model extends the cost model used in the 
multiquery optimizer, by taking differential computation into account.
The cost model used takes into
account number of seeks, amount of data read, amount of data written,
and CPU time for in-memory processing.  
Since we do not currently have a query execution engine which we 
can extend to perform differential view maintenance, we 
are unable get actual numbers.  
However, the cost model is fairly sophisticated, and
further, benefits for multiquery optimization predicted by the 
basic cost model have been verified by running rewritten queries 
on commercial database systems (\cite{rssb00:mqo}, and results 
in a companion paper on query result caching), 
giving support to the accuracy of estimated benefits.

We provide performance numbers for different percentages of updates
to the database relations; we assume that all relations are
updated by the same percentage.
To model a growing database, we have twice as many inserts as deletes.  
In our notation, a 10 percent update to a relation
consists of inserting 10\% as many tuples are currently in the 
relation, and deleting 5\% of the current tuples.

\begin{figure*}[t]
\centerline{
    \usepdf{
    \mbox{\pdfimage width 3.0in {graphs/uq4.pdf} \relax}~~
    \mbox{\pdfimage width 3.0in {graphs/qagguq4.pdf} \relax}~~
    }
    \useps{
    \psfig{file=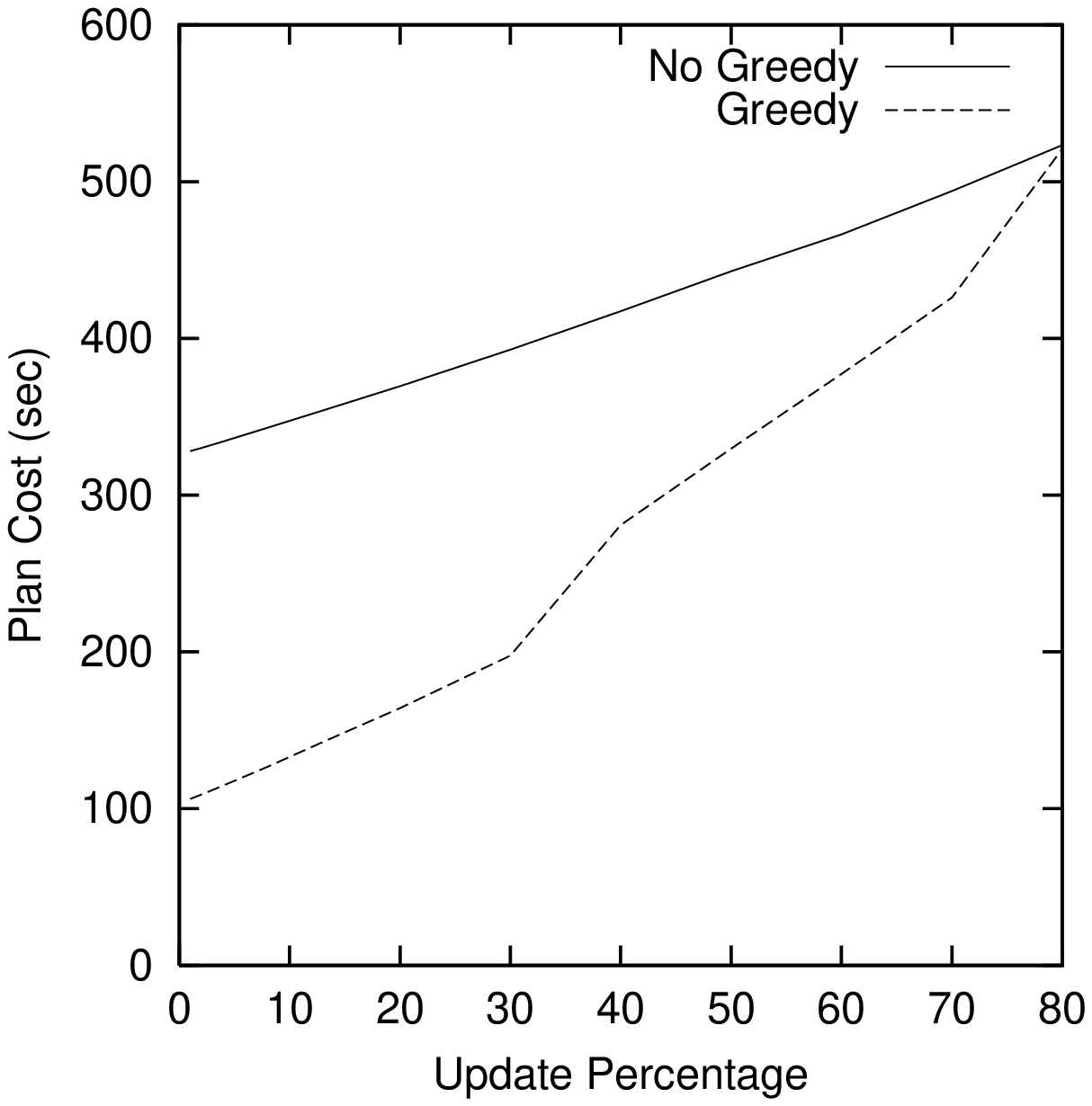,width=3.0in} ~~
    \psfig{file=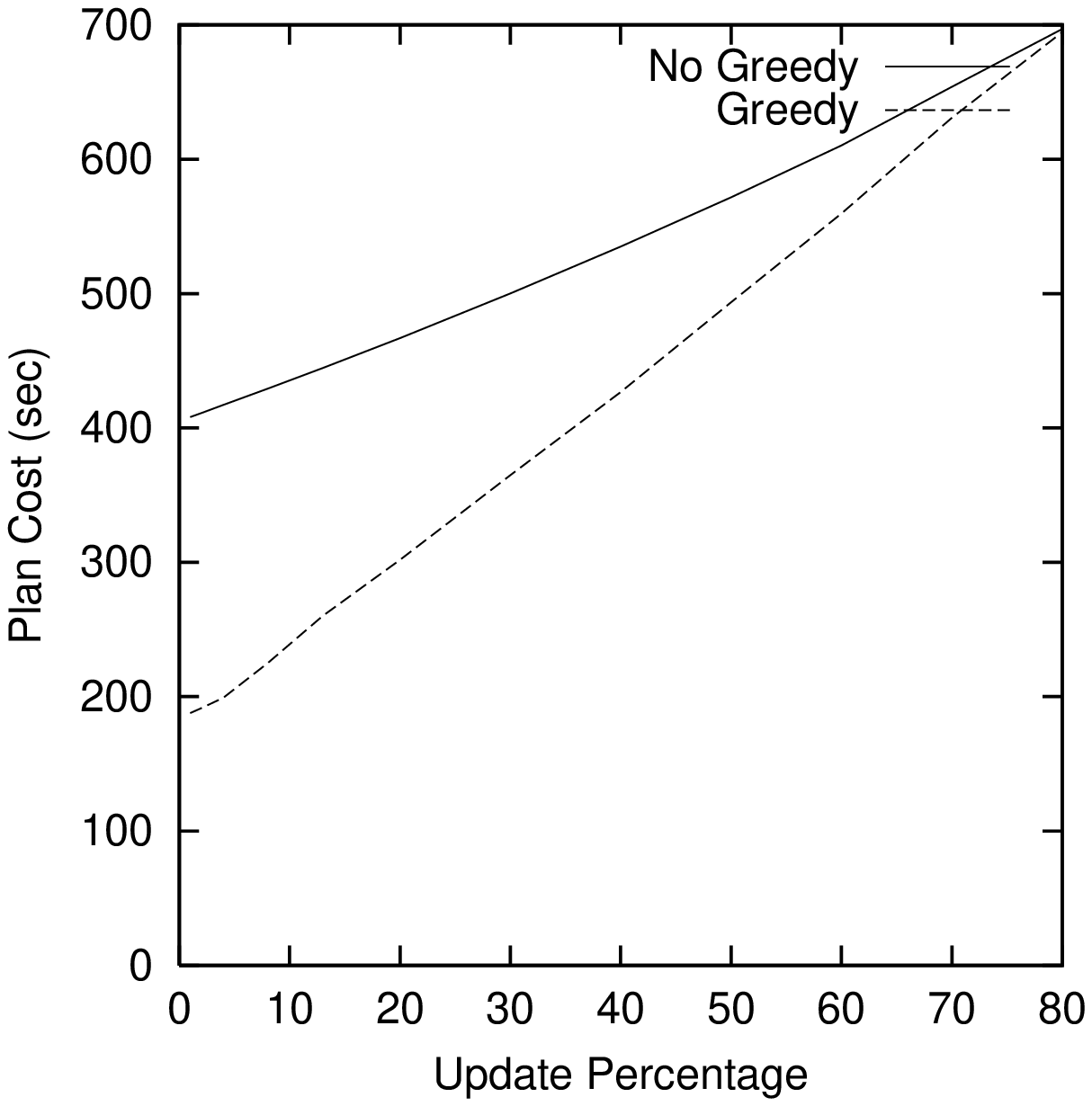,width=3.0in} ~~
    }
}
\centerline{(a) without aggregation \hspace*{2in}(b) with aggregation}
\caption{Maintaining Stand-alone Views}
\vspace*{-2ex}
\label{fig:tpcd}
\end{figure*}

We compare the performance of our greedy algorithm (referred to as
{\em Greedy} in our discussion and figures) with plain Volcano query 
optimization extended to choose between recomputation and incremental 
maintenance of views (referred to as {\em NoGreedy}).
(The algorithm of \cite{vista98:viewmaint:opt} falls in the same class
as NoGreedy, although the optimization method is somewhat different.)
For each query, we present results at different update percentages, 
ranging from 1\% to 80\%.  

The cost of view maintenance is affected by the presence of indices.
Normally, databases have indices on the primary key attributes
of each relation, to check for uniqueness.
Hence we assume that for each of the TPC-D relations, an index is
present on the primary key attributes.
However, we also ran our benchmark assuming that no indices are 
initially present, and found that all required indices got chosen
for permanent materialization by our algorithm.  
Thus, the cost of the plans we generate were not significantly
affected by the presence of indices, although the cost of
plans without our optimizations rose if indices were not already
present.

We assume a TPC-D database at scale factor
of 0.1, that is the relations occupy a total of 100 MB.  
The buffer size is set at 8000 blocks, each of size 4KB, although
we also ran some tests at a much smaller buffer size of 1000 blocks.
The tests were run on an Ultrasparc 10, with 256 MB of memory.

\vspace*{-1ex}
\subsection{Performance Results}
\vspace*{-2ex}


\noindent{\bf Maintaining Individual Views.} 
Figure~\ref{fig:tpcd} shows our results on two queries, 
the first consisting of the join of 4 relations, without aggregation, and 
the second consisting of aggregation on the same join.
As can be seen from the figures significant benefits are to be had,
especially at low update percentages, but there are benefits even at
relatively high update percentages.

\begin{figure*}
\centerline{
    \usepdf{
    \mbox{\pdfimage width 3.0in {graphs/muq1.pdf} \relax}~~
    \mbox{\pdfimage width 3.0in {graphs/muq2.pdf} \relax}~~
    }
    \useps{
    \psfig{file=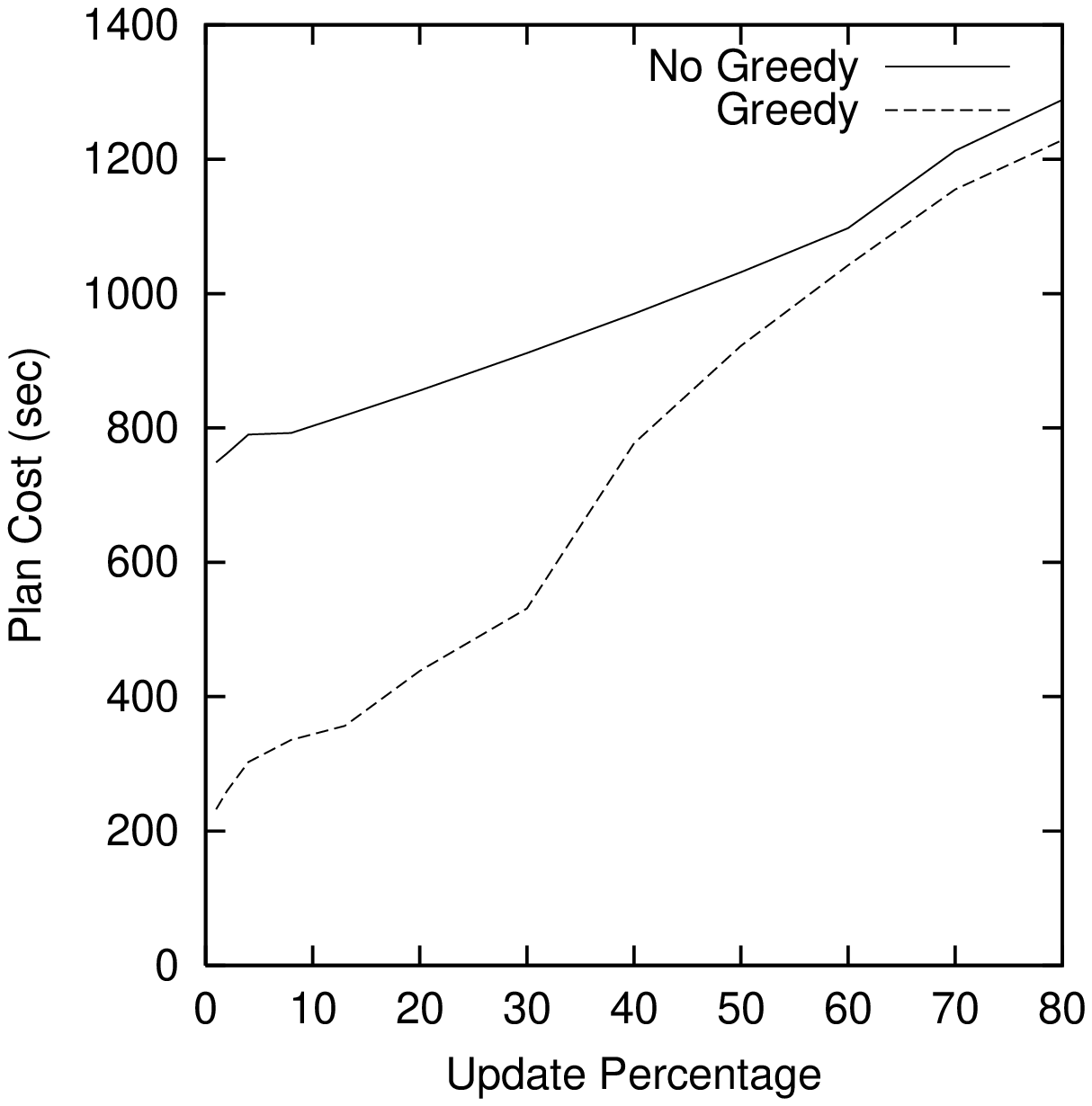,width=3.0in} ~~
    \psfig{file=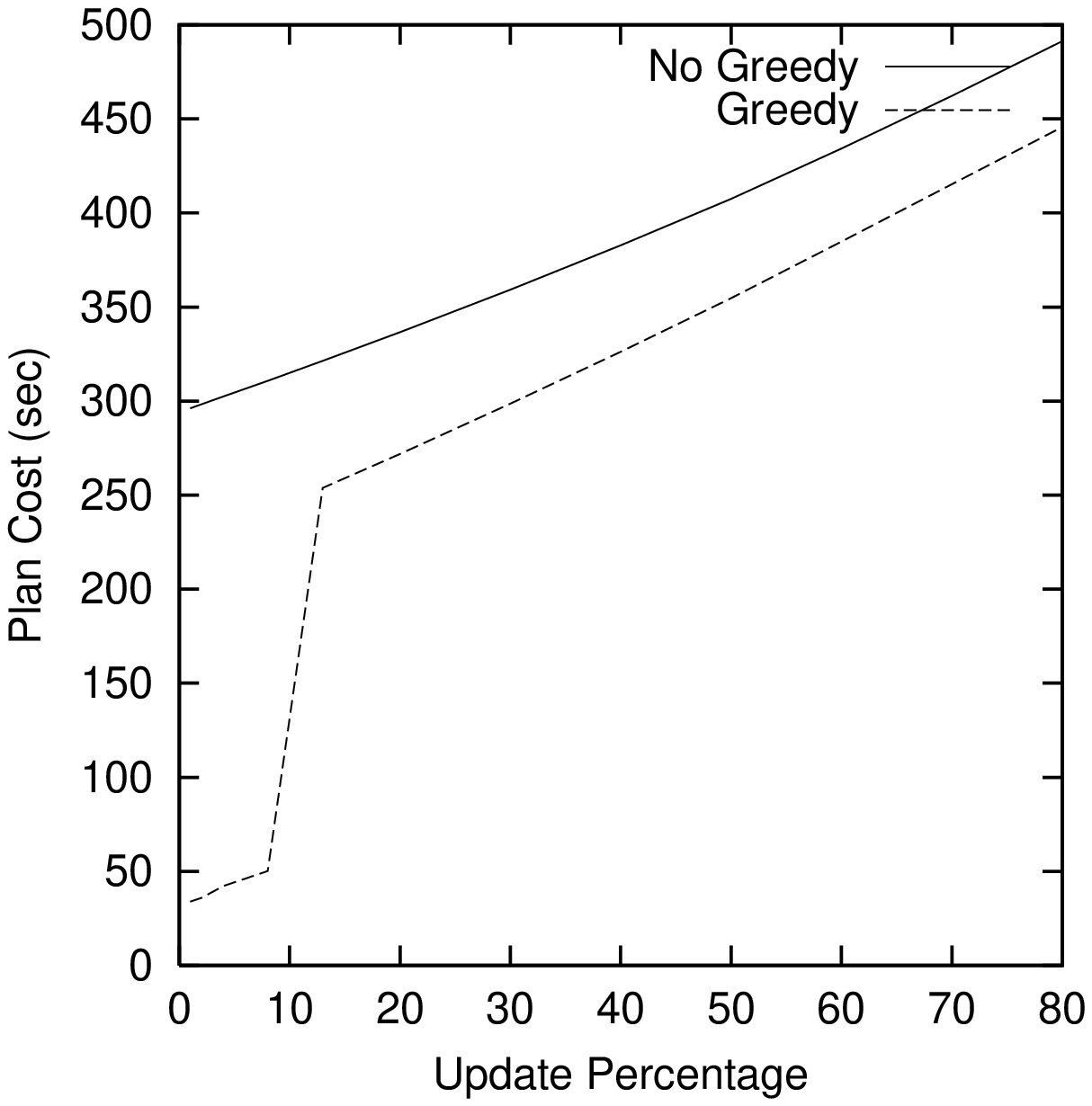,width=3.0in} ~~
    }
}
\centerline{(a) without aggregation \hspace*{1in}(b) with aggregation}
\vspace*{-2ex}
\caption{Maintaining a Set of  Views of Same Class}
\label{fig:tpcd:set}
\end{figure*}

\noindent
{\bf Maintaining a Set of Views.}
Figure~\ref{fig:tpcd:set} shows our results on two sets of queries, 
the first containing five queries without aggregation and the
second containing five queries with aggregation.
The benefit ratio due to Greedy is again excellent at lower update
percentages.    There is a jump in cost at one point, which is because
of the use of an algorithm that depends on an input fitting in memory,
and when the input does not fit in memory its cost increases sharply.

\begin{figure*}
\centerline{
    \usepdf{
    \mbox{\pdfimage width 3.0in {graphs/muq5.pdf} \relax}~~
    \mbox{\pdfimage width 3.0in {graphs/muq5ni.pdf} \relax}~~
    }
    \useps{
    \psfig{file=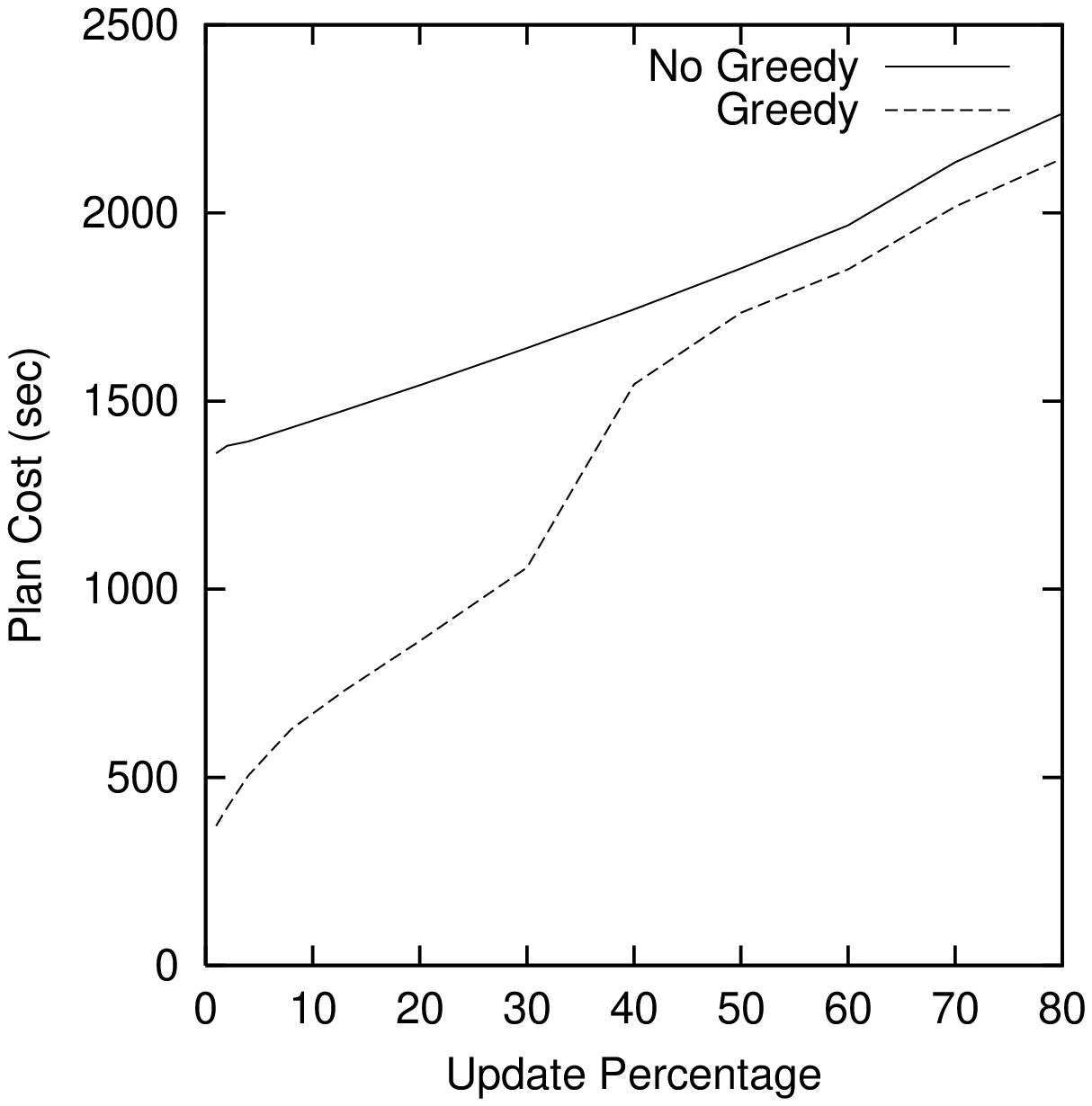,width=3.0in} ~~
    \psfig{file=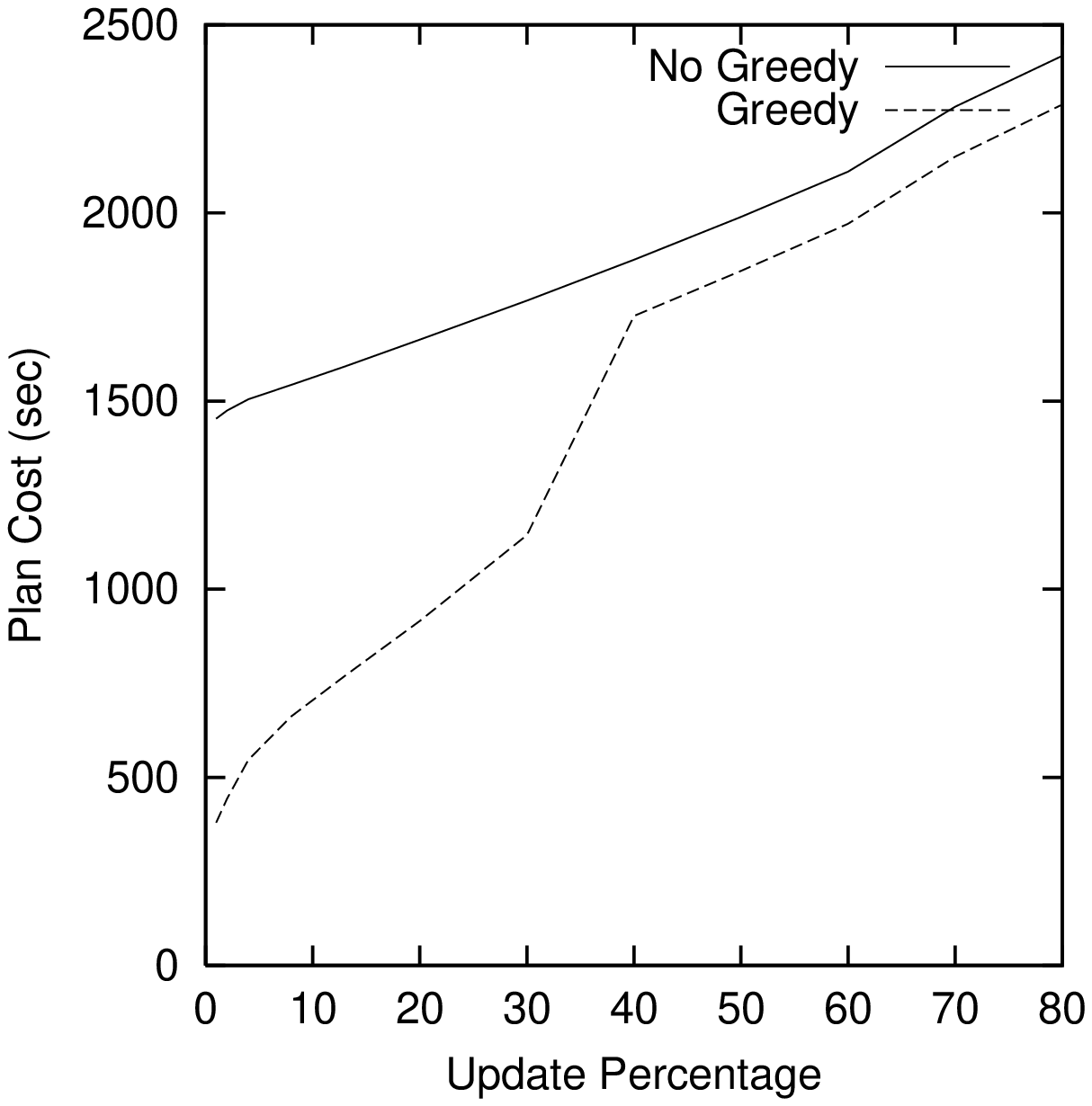,width=3.0in} ~~
    }
}
\centerline{(a) With Predefined Indices \hspace*{1in}(b) Without Predefined Indices
}
\vspace*{-2ex}
\caption{Maintaining a Large Set of  Views}
\label{fig:tpcd:largeset}
\end{figure*}

Figure~\ref{fig:tpcd:largeset} shows the results on a set of 10
queries, with indices already present on all primary key attributes,
and without any indices present initially; all required indices
got chosen for materialization.

\reminder{Perhaps a figure showing effect of no indices on many small queries}

\reminder{fill in time stats}


\noindent
{\bf Cost of Optimization.}
For a set of 10 materialized views, each a join of 3 to 4 TPC-D relations,
(whose results are shown in Figure~\ref{fig:tpcd:largeset}),
the time for Greedy optimization was 31 seconds.
Note however that 31 seconds is low compared to the savings of up to 1000
seconds obtained for one run of view maintenance, and besides it is a one-time cost 
whereas view maintenance is typically at least a daily task in 
a data warehouse.  Thus, the extra cost for our algorithms is 
worthwhile.


The number of candidates for materialization
grows exponentially with the number of relations in a query.
We are currently working on techniques to prune the set of candidates, 
in order to keep optimization time in tight control even with
a higher number of relations.

\noindent
{\bf Temporary vs. Permanent Materialization.}
Out of a total of 1600 results that were materialized
(totalling across many different queries and query sets that
we considered, and across update percentages ranging from 1 percent to
90 percent), we found that for about 1000 the recomputation cost
was less, meaning they are materialized temporarily,
and for 600 the maintenance cost was less, meaning they
were materialized permanently.
At 1 to 5 \% update rates, the ratio was 281 to 306, while at
50 to 90 \% update rates, the ratio changed to 360 to 88, in
favor of recomputation.


\reminder{ideally turn of temp/perm materialization and find
costs as compared to both materialized}

\noindent
{\bf Effect of Buffer Size.}
With a buffer size of 1000 blocks (instead of 8000 blocks),
we found that the costs of plans with and without Greedy optimization
went up, but the increase was more for recomputation plans and
the benefit ratio for small update percentages was actually
more strongly in favor of our algorithms.


%
%
%


%
%

\sections{Conclusions and Future Work}
\label{sec:concl}

The problem of finding the best way to maintain a given set of 
materialized views is an important practical problem, especially
in data warehouses/data marts, where the maintenance windows are
shrinking.
We have presented solutions that exploit commonality between different
tasks in view maintenance, to minimize the cost of maintenance.
Our techniques are easy to implement on an existing multiquery 
optimizer.
As shown by the results in section ~\ref{sec:perf}, our techniques
can generate significant speedup in view maintenance cost, 
and the increase in cost of optimization is acceptable.
We therefore believe that our results are a timely solution 
to an important practical problem.

Future work includes further heuristics to decrease optimization
cost, and implementing extensions to efficiently handle workloads 
containing queries.
We also plan to port the system to a dynamic query result
caching environment; in a companion paper, we study the issue 
of selecting results to cache dynamically, in the absence of updates.

\fullversion{
If timestamps are stored along with the tuples of the {\em delta} relations,
deferred updates can be performed at different times for different
materialized views.  
We have made the simplifying assumption that all views affected by 
the updates in the given {\em delta} relations are updated together.
We plan to extend our algorithms to allow updates to be propagated to 
different views at different times, allowing some to be updated 
more frequently. 
Query optimization in this context has to choose whether
to refresh and use an out-of-date view, or to ignore it.
}





\bibliographystyle{alpha}
\bibliography{queryopt}

\end{document}